%% file: main.tex
\documentclass[letterpaper, 10pt, twocolumn]{article}

\input{headerCommands.tex}

\begin{document}
\begin{refsection}
\input{authors.tex}

\title{Induced Directional Switching of Platicon Microcombs in Photonic Crystal Ring Resonators}

\maketitle
\thispagestyle{empty}
\noindent\textbf{\boldmath
	Microcombs in normal-dispersion photonic crystal ring resonators (PhCRs) are versatile building blocks for next-generation integrated photonic circuits, but their inherent backward-propagation bias necessitates optical circulators or complex filtering for comb extraction, creating a significant bottleneck for full on-chip integration and precluding self-injection locking schemes. In this work, we introduce Side-mode Induced Forward Forcing (SIFF), a robust method to control and reverse this directionality. By engineering auxiliary mode splittings on resonances adjacent to the pump, we steer the nonlinear dynamics to favor stable, forward-propagating platicon states. We identify an optimal coupling condition that ensures forward-comb dominance across a wide parameter range. Our findings, validated numerically and experimentally, enable circulator-free, integrated normal-dispersion microcombs compatible with self-injection locking, offering a scalable architecture for compact telecommunications and sensing systems.
}

\begin{figure*}[!t]
	\centering%
	\includegraphics[width=\textwidth]{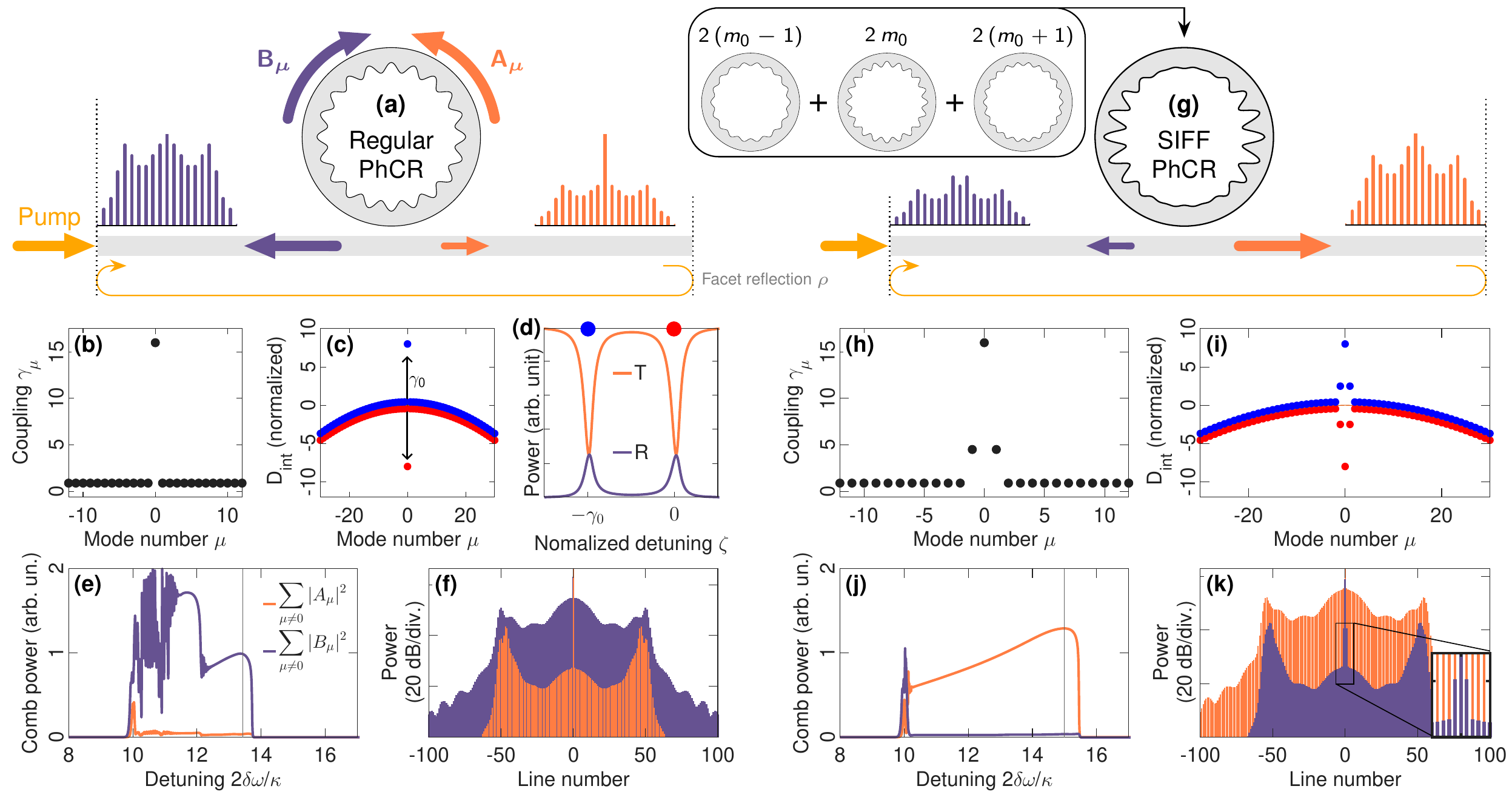}%
	\addsubcap{fig:Principle:simplePhC}{Principle of PhCR and comb directionality.}%
	\addsubcap{fig:Principle:NOSIFF:Coupling}{Spectral distribution of the bidirectional coupling factor $\gamma_\mu$ induced by a single-period PhCR.}%
	\addsubcap{fig:Principle:NOSIFF:Dint}{Overview of the normalized resonator dispersion where the mode $\mu=0$ is affected by a mode splitting ($\gamma_0 = 16$).}%
	\addsubcap{fig:Principle:NOSIFF:Splitting}{Illustration of the linear splitted resonance corresponding to the mode $\mu=0$. In our detuning convention, the red-shifted resonance is at positive detuning.}%
	\addsubcap{fig:Principle:NOSIFF:Scan}{Simulated linear detuning ramp showing the generated comb power in both direction while the pump laser scans across the red-shifted resonance. The normalized pump power is $f^2=20$.}%
	\addsubcap{fig:Principle:NOSIFF:Comb}{Simulated spectra corresponding to the scan position marked by the vertical marker in \subref*{fig:Principle:NOSIFF:Scan}.}%
	\addsubcap{fig:Principle:SIFF:PhC}{Principle of using multiple mode splitting. The PhCR is composed of superimposed corrugations with different periods such that the modes $\mu=\pm1$ are split as well.}%
	\addsubcap{fig:Principle:SIFF:Coupling}{Spectral distribution of the corresponding bidirectional coupling factor and}%
	\addsubcap{fig:Principle:SIFF:Dint}{associated normalized resonator dispersion.}%
	\addsubcap{fig:Principle:SIFF:Scan}{Simulated detuning scan (normalized pump power is $f^2=20$). Thanks to the additional splitting on the adjacent modes, the stable forward comb state takes over for most of the scan, after a brief transient initial backward comb.}%
	\addsubcap{fig:Principle:SIFF:Comb}{Simulated spectra within the main forward comb step. The backward comb lines $\pm1$ are enhanced due to the additional mode splittings (inset).}%
	\caption{%
		\textbf{Principle of comb directionality control in PhCR}
		\processCaptions
	}
	\label{fig:Principle}
\end{figure*}

\section*{Introduction}

Microresonator-based Kerr frequency combs, or microcombs, which convert continuous-wave (CW) laser light into equidistant frequency lines, have enabled the development of high-performance optical sources in compact form factors. Over the past decade, they have matured into a versatile technology for applications including miniature optical clocks, ultra-low noise microwave generation, precision ranging, and high-capacity telecommunications~\cite{Kippenberg2018,Yao2024InterdisciplinaryAdvances}.

The most established regime for microcomb generation is the dissipative Kerr soliton (DKS) state, where nonlinearity balances loss and anomalous dispersion to form self-sustained pulses~\cite{Herr2013} that emerge spontaneously as the homogeneous cavity field undergoes modulation instability. In contrast, the normal dispersion regime remains less explored due to the difficulty of accessing the required modulation instability (MI). In standard resonators, MI occurs on the unstable branch of the nonlinear resonance, typically necessitating delicate self-stabilization schemes for experimental access~\cite{Godey2014,Wang2022c}. The resulting normal-dispersion combs, often termed `platicons' or dark-pulse combs~\cite{Lobanov2015,Xue2015}, consist of switching waves connecting the upper and lower cavity field states~\cite{Coen1997,Parra-Rivas2016}. Their temporal profile varies from dark to bright pulses depending on detuning, with abrupt fronts regularized by dispersive waves appearing on the lower intensity levels~\cite{Macnaughtan2023TemporalCharacteristics,Bunel2024BroadbandKerr,Lottes2021}.

Recently, photonic crystal ring resonators (PhCRs), which incorporate a sub-wavelength grating on the resonator boundary~\cite{Arbabi2011RealizationNarrowband}, have emerged as a powerful platform for nonlinear optics~\cite{Yu2020}. The grating opens a photonic bandgap: for a ring with $2m_0$ corrugations, a bandgap occurs at the resonance fulfilling $L / \lambda_{\rm eff} = m_0$, where $L$ is the roundtrip length. This manifests as a mode splitting that acts as a localized dispersion defect, enabling phase-matched modulation instability for optical parametric oscillation (OPO) and platicon formation in the normal dispersion regime~\cite{Stone2023WavelengthaccurateNonlinear, Black2022, Lobanov2015, Yu2021}. While other techniques for mode-shifting, such as avoided mode crossings~\cite{Xue2015}, coupled resonators~\cite{Helgason2020}, or nonlinear shifts from counter-propagating pumps~\cite{Yang2024CrossphaseModulation}, have been explored, PhCRs offer a deterministic and mode-specific approach.

Physically, the photonic crystal effect acts as a frequency-dependent directional coupling. The corrugation functions as a Bragg mirror, coupling the two inherently degenerate counter-propagating directions (see \cref{fig:Principle:simplePhC}). This creates two supermodes (standing waves), which are frequency up-shifted (blue) and down-shifted (red), appearing as split resonances in transmission (see \cref{fig:Principle:NOSIFF:Dint}). 
A significant associated caveat of PhCRs is the resulting bidirectional comb dynamics. When only the pumped mode is split, comb initialization occurs almost exclusively in the backward direction (opposite to the pump) as shown in \cref{fig:Principle:simplePhC,fig:Principle:NOSIFF:Comb}~\cite{Lucas2021DarkPulseDynamics}, a direct consequence of pumping the red-shifted supermode where the intracavity power is predominantly backward-propagating.

While this backward bias can be enhanced via bus waveguide reflectors to boost backward comb efficiency~\cite{Zang2025LaserPower}, it poses fundamental extraction challenges. Effective solutions require either bulky, non-integrated optical circulators or complex phase-shifted Bragg gratings to route comb components away from the pump. Drop ports are ineffective as light is still extracted backward at the injection point.
Moreover, while the increased pump back-reflection in PhCR has been exploited in anomalous dispersion for soliton generation via self-injection locking (SIL)~\cite{Ulanov2024SyntheticReflection}, this approach relies on inherently forward-propagating solitons and is incompatible with backward-propagating platicons which could destabilize the driving laser in the absence of precise filtering. Although stochastic backward-to-forward switching has been observed at higher detuning and power~\cite{Lucas2021DarkPulseDynamics,Lobanov2025ControllableSwitchingPlaticon}, this process remains unreliable and lacks practical stability. Forward propagation is therefore essential for both efficient on-chip routing and laser stability, motivating the need for a robust method to enforce forward operation in PhCRs.

In this work, we demonstrate that introducing additional engineered mode splittings on the modes adjacent to the pump, as illustrated in \cref{fig:Principle:SIFF:PhC}, significantly expands the parameter range for stable forward-propagating combs. We numerically investigate the conditions for this Side-mode Induced Forward Forcing (SIFF) and experimentally validate the reversal of comb directionality. The SIFF method directly engineers the resonator to favor forward propagation at the source, offering a more fundamental solution.

\section*{Results}

\subsection{Principle and modeling}

The principle of directional control in PhCRs is illustrated in \cref{fig:Principle} through numerical analysis. We first consider a standard PhCR with a single corrugation period. In the normal dispersion regime, the resonance frequencies exhibit a negative parabolic mismatch relative to an equidistant grid, expressed as $D_{\rm int}(\mu) = (\omega_\mu - \omega_0) - D_1 \mu = \frac{D_2}{2} \mu^2$, where $\mu$ is the mode index relative to the pumped resonance ($\omega_0$), $D_1/2\pi$ is the free spectral range (FSR), and $D_2 < 0$ is the group velocity dispersion (GVD) parameter (\cref{fig:Principle:NOSIFF:Dint}), and higher order dispersion is neglected. The integrated photonic crystal induces a spectrally-selective Bragg reflection, resulting in a localized mode splitting at $\mu=0$ that manifests as distinct blue- and red-shifted supermodes.

To capture the nonlinear dynamics, we employ a coupled system of bidirectional Lugiato-Lefever equations (LLEs)~\cite{Lugiato1987,Haelterman1992} describing the intra-cavity field evolution in the forward ($\tilde{A}_\mu$) and backward ($\tilde{B}_\mu$) directions. In the spectral domain, these equations incorporate the mode-dependent coupling rate $\gamma_\mu$ induced by the grating, which facilitates energy transfer between counter-propagating directions. The normalized equations are given by~\cite{Skryabin2020}:
\begin{equation} \label{eq:lles}
	\begin{aligned}
		\dfrac{d}{d t} \begin{bmatrix}
			\tilde{A}_\mu \\
			\tilde{B}_\mu
		\end{bmatrix} = &
		\begin{bmatrix}
			\tilde{\xi}(B)  & i \, \dfrac{\gamma_\mu}{2}\\
			i \, \dfrac{\gamma_\mu}{2} & \tilde{\xi}(A)
		\end{bmatrix}
		\begin{bmatrix}
			\tilde{A}_\mu \\
			\tilde{B}_\mu
		\end{bmatrix}
		+   i \,  \begin{bmatrix}
			\mathcal{F}\left[ |A|^2 \, A \right]_\mu \\
			\mathcal{F}\left[ |B|^2 \, B \right]_\mu
		\end{bmatrix} \\
		&+ f \begin{bmatrix}
			\sqrt{1-\rho} \\
			\sqrt{\rho} e^{i\phi}
		\end{bmatrix}
		\delta_{\mu=0}
	\end{aligned}
\end{equation}
where the slow time $t$ is normalized to twice the photon lifetime $2/\kappa$ ($\kappa$ being the resonance linewidth). Here, $\mathcal{F}$ denotes the Fourier series operator, $\tilde{A}_\mu$ and $\tilde{B}_\mu$ are normalized field amplitudes, and $f = \sqrt{8 g_0\kappa_{\rm ex} P_{\rm in}/\hbar\omega_0\kappa^3}$ is the pump field strength~\cite{Herr2013}. The pump detuning is defined as $\zeta = 2(\omega_0 - \Gamma_0/2 - \omega_{\rm pump})/\kappa$, while $d_2 = D_2/\kappa$ represents the normalized GVD. The operator $\tilde{\xi}(A) = - 1 + i \left(-\zeta + 2 \sum_{\mu'} |\tilde{A}_{\mu'}|^2 + d_2 \mu^2 \right)$ accounts for losses, detuning, self- and cross-phase modulation, and dispersion. The normalized linear coupling rates are $\gamma_\mu = 2 \Gamma_\mu / \kappa$.
The bidirectional coupling is further influenced by parasitic back-reflections from the chip facets, with $\rho$ representing the facet reflectivity and $\phi$ the associated back-reflection phase. For the initial modeling of the SIFF mechanism, we assume an ideal case with no facet reflection ($\rho = 0$).
The equations are solved using the split-step Fourier method.

\begin{figure*}[!t]
	\centering%
	\includegraphics[width=\textwidth]{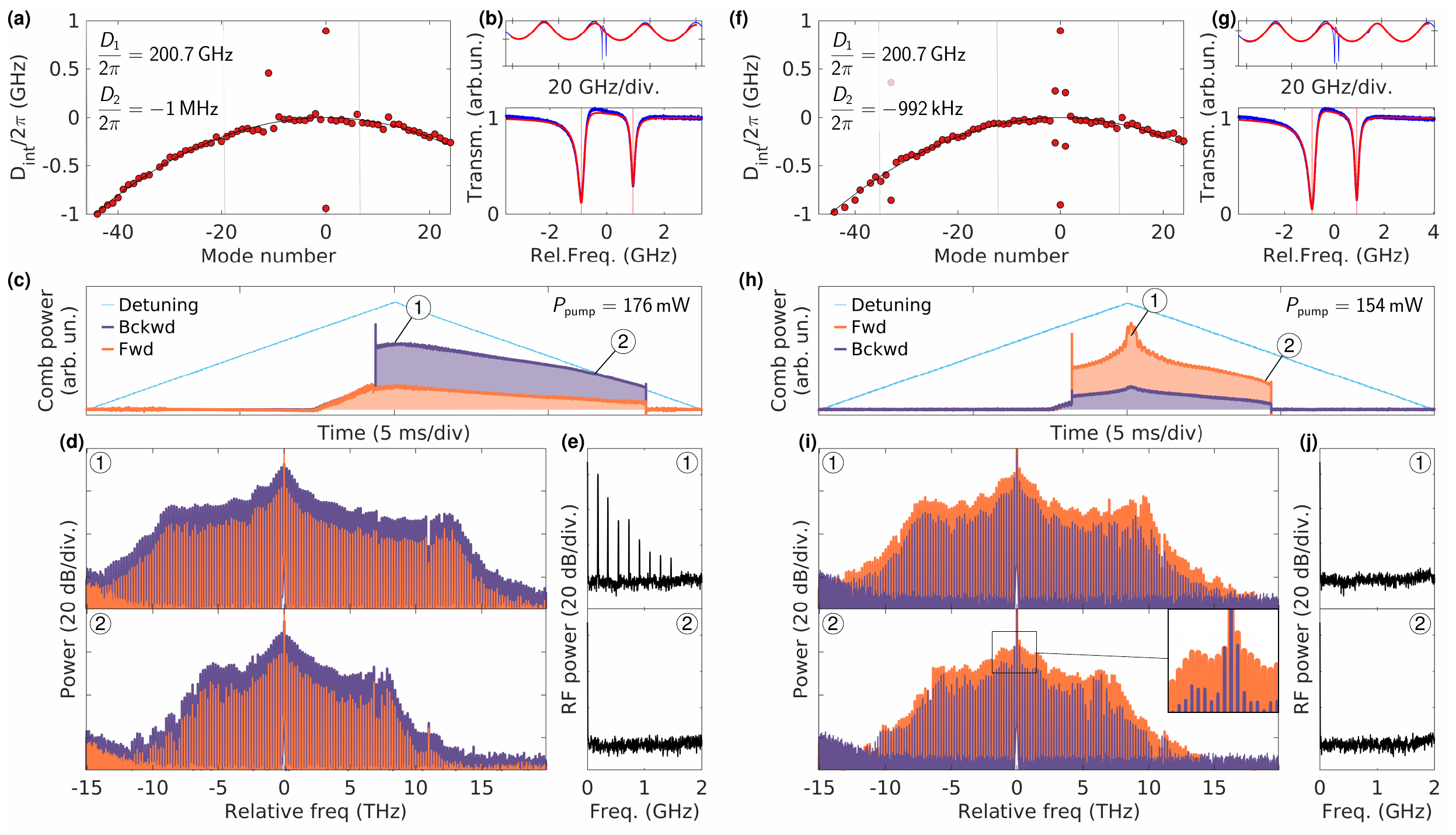}%
	\addsubcap{fig:Exp:SimplePhC:Dint}{Measured integrated dispersion of a regular PhCR, showing the measured resonance frequencies deviation from a fixed FSR spacing as a function of the mode number.}%
	\addsubcap{fig:Exp:SimplePhC:PumpRes}{Fitting of the split resonance for $\mu = 0$. The top panel shows the raw transmission with the chip facets Fabry Pérot (FP) oscillations. The bottom panel focuses on the split resonance after normalization by the FP.}%
	\addsubcap{fig:Exp:SimplePhC:ScanScope}{Oscilloscope trace showing the evolution of the generated comb power as the detuning is scanned back and forth across resonance the red side of the mode splitting.}%
	\addsubcap{fig:Exp:SimplePhC:Spectra}{Optical spectra recorded in the forward and backward direction, corresponding to the states 1 (high detuning) and 2 (low detuning) marked in the scan in \subref*{fig:Exp:SimplePhC:ScanScope}.}%
	\addsubcap{fig:Exp:SimplePhC:Noise}{Intensity noise of the comb states 1 and 2, recorded on a electronic spectrum analyzer.}%
	\addsubcap{fig:Exp:SIFF:Dint}{Measured integrated dispersion of a SIFF PhCR.}%
	\addsubcap{fig:Exp:SIFF:PumpRes}{Split resonance for $\mu = 0$. The overall lineshape and resonance position relative to the chip FP was selected to be similar to the regular device.}%
	\addsubcap{fig:Exp:SIFF:ScanScope}{Experimentally-measured evolution of the generated comb light in both directions during a detuning scan.}%
	\addsubcap{fig:Exp:SIFF:Spectra}{Measured optical spectra in the forward and backward directions corresponding to the high detuning state (1) and low detuning state (2) marked in \subref*{fig:Exp:SIFF:ScanScope}. The inset focuses on the enhanced lines $\pm 1$ linked to side mode splitting.}%
	\addsubcap{fig:Exp:SIFF:Noise}{Spectrum of the intensity noise of the combs shown in  \subref*{fig:Exp:SIFF:Spectra}.}%
	\caption{%
		\textbf{Experimental demonstration of PhCR comb `forward forcing'.}
		\processCaptions
	}
	\label{fig:exp:compare}
\end{figure*}

\begin{figure}[t]
	\centering%
	\includegraphics[width=\linewidth]{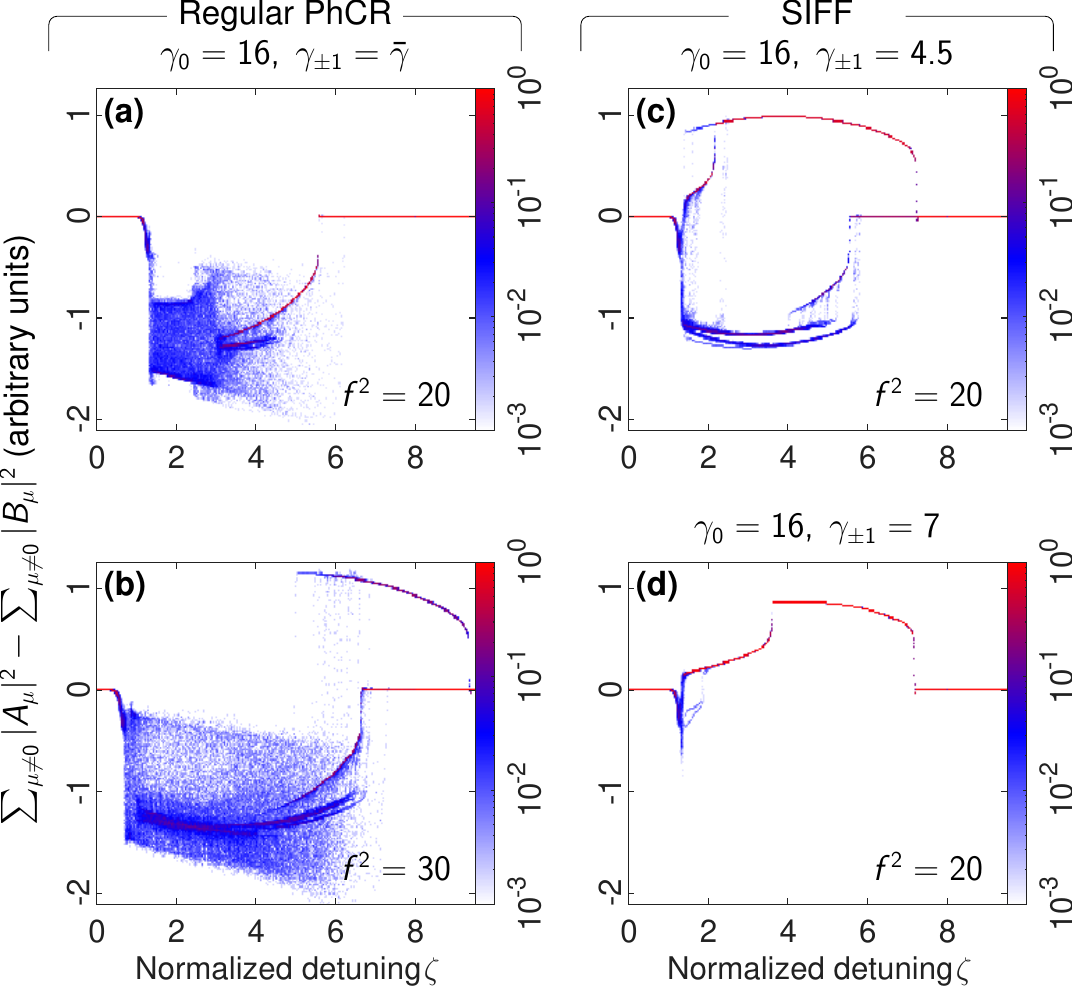}%
	\addsubcap{fig:scanStats:noSIFFLow}{Regular PhCR (same configuration as \cref{fig:Principle}).}%
	\addsubcap{fig:scanStats:noSIFFHigh}{Regular PhCR at $f^2 = 30$.}%
	\addsubcap{fig:scanStats:SIFF}{Moderate SIFF ($\gamma_{\pm 1} = 4.5$, same as \cref{fig:Principle}) at $f^2 = 20$.}%
	\addsubcap{fig:scanStats:SIFFoptim}{Optimal SIFF ($\gamma_{\pm 1} = 7$) at $f^2 = 20$.}%
	\caption{%
		\textbf{Probability maps of comb direction during detuning scans.}
		Probability distribution maps of the difference between forward and backward comb power during detuning scans. Negative values indicate dominant backward propagation, while positive values indicate forward dominance. The colorbar represents the probability density, analogous to an oscilloscope in persistence mode.
		\processCaptions
	}
	\label{fig:scanStats}
\end{figure}

The nonlinear dynamics of a regular PhCR ($\gamma_0 = 16$) are illustrated in \cref{fig:Principle:NOSIFF:Scan,fig:Principle:NOSIFF:Comb}. To account for intrinsic Rayleigh scattering observed in our devices, a baseline coupling of $\bar{\gamma} = 0.88$ is included for all modes. Upon scanning the pump laser across the red-shifted supermode at $f^2 = 20$, the microcomb initiates and remains predominantly in the backward direction. This evolution begins with a chaotic regime at low detunings, characterized by marked power fluctuations, followed by a transition into a stable platicon `step'. By the end of this plateau, the intracavity power resides almost exclusively in the backward-propagating field (\cref{fig:Principle:NOSIFF:Comb}).

Next, we introduce additional mode splittings at $\mu = \pm 1$ using the superposition of multiple grating periods~\cite{Lu2020, Lucas2023TailoringMicrocombs}, as shown in \cref{fig:Principle:SIFF:PhC}. By setting $\gamma_{\pm 1} = 4.5$ (\cref{fig:Principle:SIFF:Coupling}), the resulting directional coupling $\gamma_\mu$ and  $D_{\rm int}$ profile are modified (\cref{fig:Principle:SIFF:Dint}). Simulations of the detuning scan for this configuration reveal a dramatic change in behavior (\cref{fig:Principle:SIFF:Scan}). Although the comb undergoes a brief backward-propagating transient, the dynamics rapidly switch to a stable, forward-propagating platicon state. The resulting comb step is notably smoother and regular, and the final state is dominated by the forward direction (\cref{fig:Principle:SIFF:Comb}).

These numerical results demonstrate that the targeted introduction of mode splittings on adjacent resonances can deterministically reverse the microcomb's dominant propagation direction, which we dub SIFF.
This phenomenon suggests a mechanism where the engineered coupling at $\mu = \pm 1$ acts as a coherent seed. While the comb initially forms in the backward direction, the side-mode splitting induces a forward-propagating modulation that steers the nonlinear dynamics toward a forward-propagating platicon state. Once established, this forward pulse more efficiently utilizes the externally injected pump power. Due to the cross-phase modulation and the resulting competition for the intracavity pump field, the growth of the forward state depletes the available energy, effectively suppressing the backward-propagating pulse and preventing the coexistence of the two counter-propagating combs.
The specific conditions and stability criteria for this transition are investigated in the following sections.

\begin{figure}[t]
	\centering%
	\includegraphics[width=0.95\linewidth]{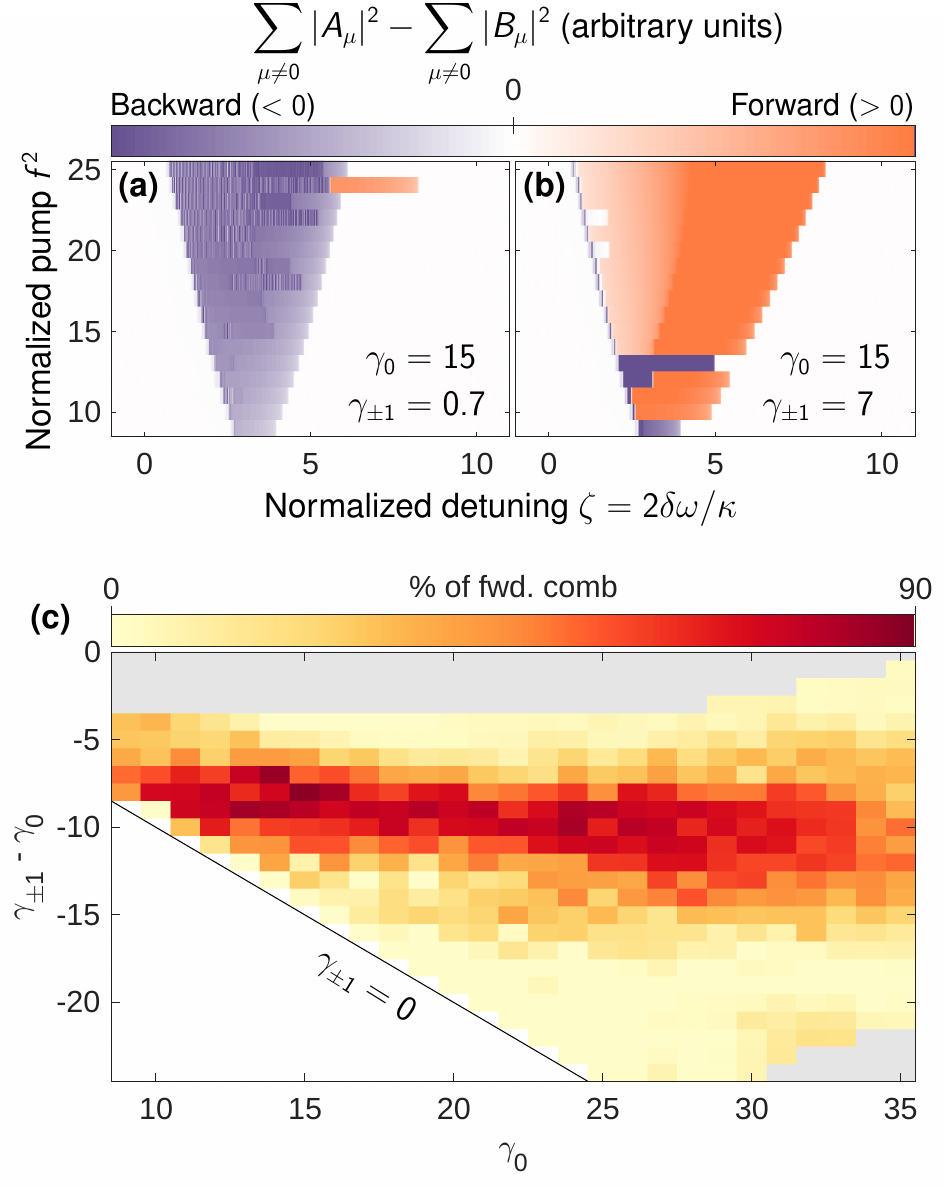}%
	\addsubcap{fig:sidemodeMaps:noSIFFMap}{Phase diagram showing the dominant comb direction at different values of pump power and laser detuning, in the absence of SIFF and a pump mode splitting of $\gamma_0 = 15$.}%
	\addsubcap{fig:sidemodeMaps:SIFFMap}{Same diagram obtained when the modes adjacent to the pump are split by $\gamma_{\pm 1} = 7$}%
	\addsubcap{fig:sidemodeMaps:SIFFOptimMap}{Diagram showing the percentage of dominant forward comb state aggregated over the $(\zeta, F^2)$ maps for different values of $\gamma_0$ and $\gamma_{\pm 1}$.}%
	\caption{%
		\textbf{SIFF optimization}
		\processCaptions
	}
	\label{fig:sidemodeMaps}
\end{figure}

\begin{figure*}[t]
	\centering%
	\includegraphics[width=\textwidth]{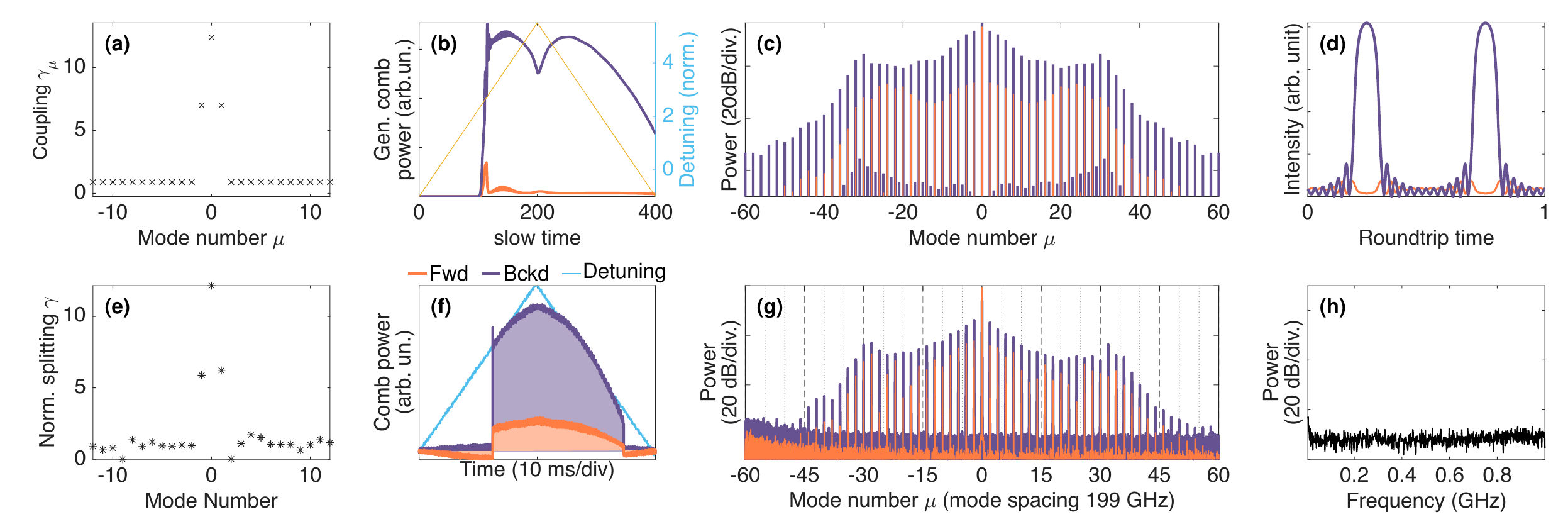}%
	\addsubcap{fig:DualPlat:Sim:gamma}{Synthetic coupling profile assumed for the simulations $\gamma_0=12.4$, $\gamma_{\pm 1} = 7$.}%
	\addsubcap{fig:DualPlat:Sim:Scan}{Simulated detuning scan for a normalized pump power $f^2 = 16$, showing the evolution of the comb power in both directions.}%
	\addsubcap{fig:DualPlat:Sim:Spectra}{Optical spectrum of the forward and backward combs at $t=300$ in the simulated scan. The odd line numbers are strongly supressed.}%
	\addsubcap{fig:DualPlat:Sim:Temporal}{Temporal profile corresponding to the combs shown in \subref*{fig:DualPlat:Sim:Spectra}}%
	\addsubcap{fig:DualPlat:Exp:gamma}{Experimentally measured mode splitting distribution in a sub-optimal SIFF PhCR.}%
	\addsubcap{fig:DualPlat:Exp:Scan}{Experimentally-measured evolution of the generated comb light in both directions during a detuning scan.}%
	\addsubcap{fig:DualPlat:Exp:Spectra}{Measured optical spectra of the forward and backward combs near the maximum detuning of the scan. Every odd mode comb line is suppressed.}%
	\addsubcap{fig:DualPlat:Exp:Noise}{spectrum of the intensity noise of the corresponding combs.}%
	\caption{%
		\textbf{\boldmath Platicon crystal formation at larger $\gamma_{\pm 1}$}
		\processCaptions
	}
	\label{fig:exp:DualPlat}
\end{figure*}

\subsection{Experimental demonstration}

To experimentally validate the SIFF effect, we fabricated two devices: a regular PhCR with a single corrugation affecting only the pumped mode (\cref{fig:exp:compare}a-e), and a second device incorporating two additional mode splittings at $\mu = \pm 1$ via the superposition principle (\cref{fig:exp:compare}f-j). Both microring resonators were designed with identical geometries and fabricated using a \qty{570}{\nm} thick tantalum pentoxide (\ce{Ta2O5}) layer on silica without top cladding~\cite{Jung2020}. The spectral characteristics were characterized using a frequency-calibrated laser scan~\cite{Fujii2020}.

The resonators have a radius of \qty{109}{\micro\meter}, yielding a free spectral range (FSR) of $D_1/2\pi = \qty{200.7}{\GHz}$ and a normal dispersion coefficient of $D_2/2\pi \approx \qty{-1}{\MHz}$. The measured resonance linewidth was $\kappa/2\pi \approx \qty{230}{\MHz}$ in the slightly overcoupled regime, with an estimated intrinsic quality factor of \num{2.4e6}. The standard PhCR features a single-period corrugation ($2\,m_0 = 1626$ periods) with an \qty{11}{\nm} peak-to-peak amplitude, resulting in a pump-mode splitting of $\Gamma_0/2\pi = \qty{1.82}{\GHz}$ ($\gamma_0 = 15.8$). The SIFF device employs the same primary corrugation supplemented by two additional gratings ($\pm 2$ periods relative to the pump) with amplitudes \qty{30}{\percent} of the main period (\qty{3.3}{\nm}). This yielded measured splittings of $\Gamma_0/2\pi = \qty{1.82}{\GHz}$, $\Gamma_{-1}/2\pi = \qty{553}{\MHz}$, and $\Gamma_1/2\pi = \qty{571}{\MHz}$ ($\gamma_{\pm 1} \approx 4.88$).

A parasitic \qty{10}{\percent} reflection from the etched chip facets, consistent with the Fresnel reflection at the waveguide-air interface (see \cref{fig:Principle:simplePhC,fig:Principle:SIFF:PhC}), creates a Fabry-Pérot (FP) cavity within the chip. The interaction between this FP background and the microring resonances results in Fano-like lineshapes~\cite{Fan2002}. To isolate the SIFF effect from the influence of back-reflection phase, we selected devices with near identical relative detunings and lineshapes (\cref{fig:Exp:SimplePhC:PumpRes,fig:Exp:SIFF:PumpRes}). A detailed numerical investigation of the back-reflection phase is provided in the Supplemental Material.
Notably, this back-reflection also restricts the maximum observable contrast between forward and backward comb powers to \qty{10}{\decibel}, whereas simulations predict an extinction exceeding \qty{20}{\decibel}.

The experimental setup for comb generation utilizes an amplified tunable CW laser at \qty{1550}{\nm} followed by a bandpass filter for ASE rejection. A fiber circulator is used to inject the pump while simultaneously collecting the backward-propagating comb and any back-reflected light via lensed fibers. The forward-propagating light is collected from the output facet. Both directions are monitored using an optical spectrum analyzer (OSA), and the generated comb power is measured on photodiodes after filtering the pump line with an optical notch filter. Insertion losses were carefully calibrated to ensure accurate power imbalance measurements.

Upon scanning the pump laser into the nonlinear regime, we observed two distinct behaviors. In the regular device ($P_{\rm in} \sim \qty{176}{\mW}$), the microcomb emerges as the laser scans across the red-shifted supermode at relatively large detuning (\cref{fig:Exp:SimplePhC:ScanScope}). After a transient phase of primary sideband generation, a stable comb forms, predominantly in the backward direction. This state exhibits clear hysteresis: once initiated, the comb persists over a wide range of decreasing detunings. Optical spectra acquired at both ends of the platicon step exhibit characteristic dark-pulse features (\cref{fig:Exp:SimplePhC:Spectra}). At high detuning, the state exhibits breathing dynamics, evidenced by distinct modulation tones in the RF intensity noise (\cref{fig:Exp:SimplePhC:Noise})~\cite{Lucas2017b}, whereas at lower detunings, the comb transitions to a narrower, low-noise state.

We note that this breathing trend is inverted relative to our idealized simulations. This discrepancy likely arises from the complex interplay between parasitic back-reflections and local avoided mode crossings present in the physical device, which modify the stability boundaries. Nevertheless, this deviation does not alter our primary conclusion: at this power level, the comb is systematically backward. Increasing the pump power to \qty{\sim 230}{\mW} enables rare transitions of the comb from an unstable backward state to a stable forward state~\cite{Lucas2021DarkPulseDynamics}, similar to the simulations shown in \cref{fig:scanStats:noSIFFHigh}. Such transitions occur in only \qty{\sim 8}{\percent} of cases (measured over 25 consecutive scans).

In striking contrast, the SIFF device yields a microcomb that is predominantly forward-propagating (\cref{fig:Exp:SIFF:ScanScope}, $P_{\rm in} \sim \qty{154}{\mW}$). While the hysteresis behavior remains similar, the forward-propagating state dominates the entire existence range of the platicon. This state is confirmed to be low-noise across its stability range (\cref{fig:Exp:SIFF:Spectra,fig:Exp:SIFF:Noise}). As expected, the backward power in the $\mu = \pm 1$ lines is enhanced in the SIFF design due to the engineered coupling (see inset in \cref{fig:Exp:SIFF:Spectra}). In this configuration, we measured forward propagation in \qty{\sim 68}{\percent} of scans, with the comb remaining backward in the remaining cases. These results confirm that SIFF provides a robust mechanism for directional control of platicon microcombs in normal-dispersion PhCRs.

\subsection{Systematic analysis and SIFF optimization}

To quantify the reliability of forward comb generation, we performed statistical analysis by simulating 100 detuning scans for each configuration and calculating the probability distribution of the forward -- backward power difference.
\Cref{fig:scanStats} presents probability maps of the comb direction (where negative values indicate backward dominance and positive values indicate forward dominance) as a function of detuning.
For a simple PhC at $f^2=20$ (\cref{fig:scanStats:noSIFFLow}), the comb remains 100\% backward with a distribution spread indicating breathing and instabilities.
At higher power $f^2=30$ (\cref{fig:scanStats:noSIFFHigh}), while most scans still produce purely backward combs with instabilities, approximately 28\% of scans exhibit direction switching to forward comb, consistent with previous observations in regular PhCRs~\cite{Lucas2021DarkPulseDynamics,Lobanov2025ControllableSwitchingPlaticon}.
Introducing moderate SIFF with $\gamma_{\pm 1}=4.5$ at $f^2=20$ (\cref{fig:scanStats:SIFF}) results in forward comb generation as the main outcome in 70\% of scans, with the remaining 30\% producing unstable backward combs.
With optimal SIFF parameters $\gamma_{\pm 1}=7$ (\cref{fig:scanStats:SIFFoptim}), 100\% of scans produce stable forward combs with no instabilities.

To establish the optimal regime of the SIFF effect, we performed a systematic numerical study over a wide parameter space. Starting from a zero cavity field, we simulated detuning scans ($\zeta$) for varying pump power levels ($f^2$). \Cref{fig:sidemodeMaps:noSIFFMap} maps the power imbalance between the forward and backward directions for a regular device ($\gamma_0 = 15, \gamma_{\pm 1} = 0.7$). In this configuration, the backward-propagating comb remains dominant throughout nearly the entire existence range of the platicon.

In contrast, the introduction of optimized side-mode splitting ($\gamma_{\pm 1} = 7$) induces a radical shift in the nonlinear dynamics (\cref{fig:sidemodeMaps:SIFFMap}), with the forward-propagating state dominating the full comb existence range.
Interestingly, the simulations reveal that the comb still initiates as a brief backward-propagating burst before switching to the forward direction.
At low detunings, an intermediate regime emerges where both propagation directions exhibit comparable power levels with a forward bias, prior to the forward state fully dominating. This is a genuine detuning-dependent state, not a scan-speed artifact, where most power flows forward while the backward comb retains measurable energy. The detuning span of this regime corresponds to the instability region of a pure backward comb in a standard device. Notably, the SIFF coexistence state remains stable, and the coexistence area increases with larger pump mode splitting.

To identify the optimal conditions for SIFF, we computed these maps for a range of pump splittings $\gamma_0 \in [10, 35]$ and side-mode splittings $\gamma_{\pm 1}$.
We define the ``SIFF probability'' as the ratio between the parameter space area (in detuning and power) where the forward comb is dominant and the total area where any stable comb state (forward or backward) exists.
This normalized metric is summarized in the optimization map in \cref{fig:sidemodeMaps:SIFFOptimMap}.
A clear ``valley'' of optimal SIFF operation emerges, following a linear relationship defined by a nearly constant offset: $\gamma_0 - \gamma_{\pm 1} \approx 9$.
However, due to fabrication variability, we were unable to realize a device with parameters in this optimized SIFF region.

Physically, this behavior arises from optimized phase-matching between the pumped mode and the adjacent split modes. The enhanced field generation at $\mu = \pm 1$ triggers coherent forward re-injection, seeding a forward-propagating platicon that suppresses the initial backward state. This mechanism is resilient to pump power variations (see \cref{fig:sidemodeMaps:SIFFMap}) and remains effective across more than one decade of dispersion change (see Supplementary Information).

Conversely, when $\gamma_{\pm 1}$ is too large (typically $\gamma_{\pm 1} \gtrsim \gamma_0 - 6$), the SIFF effect is suppressed. As illustrated in \cref{fig:exp:DualPlat} for $\gamma_0 = 12.4$ and $\gamma_{\pm 1} = 7$, the system reverts to a dominant backward-propagating state. In this ``over-split'' regime, the system preferentially generates a dual platicon crystal, comprising two equidistant pulses within the cavity (\cref{fig:DualPlat:Sim:Temporal}). The resulting spectrum (\cref{fig:DualPlat:Sim:Spectra}) exhibits the characteristic signatures of harmonic mode-locking, where every odd comb line is suppressed, effectively doubling the repetition rate. This behavior is explained by the excessive frequency shift of the side modes, which inhibits phase-matching. Consequently, the $\mu = \pm 1$ lines are suppressed, preventing the forward-seeding mechanism and leaving the inherent backward bias of the PhCR unchallenged. This regime was experimentally confirmed in a device with $\gamma_0 = 12.3$ and $\gamma_{\pm 1} = 6$ (\cref{fig:DualPlat:Exp:gamma}), where the measured spectrum and directional bias perfectly match the harmonic platicon crystal predicted by our simulations.

Beyond the idealized case, we highlight the critical impact of parasitic back-reflections on the nonlinear dynamics. Even a relatively small reflection (10\%) from the chip facets can drastically alter the comb stability, as detailed in the Supplementary Information. Crucially, our results demonstrate that the SIFF mechanism provides significant leverage, extending the forward comb's existence range compared to regular devices even in the presence of these experimental imperfections. This underscores the robustness of our approach and the necessity of managing facet reflections to fully exploit the directional control offered by SIFF.

\section*{Discussion}

In summary, we have introduced and demonstrated the Side-mode Induced Forward Forcing effect as a robust method to control the propagation direction of platicon microcombs in photonic crystal ring resonators. By strategically engineering mode splittings on resonances adjacent to the pump, we successfully reversed the inherent backward bias of regular PhCRs, enabling the generation of stable, low-noise forward-propagating combs.

Although backward reflection can also be suppressed by splitting side modes alone while leaving the pump mode unsplit~\cite{Jin2025BandgapdetunedExcitation}, this approach forfeits the ability to tailor the platicon state pattern and comb envelope via pump-resonance splitting~\cite{Yu2021}. Our SIFF strategy overcomes this limitation, leveraging split-pump dynamics to maintain full design flexibility while enforcing unidirectionality through side-mode engineering.

Looking forward, several avenues remain to be explored to mature this technology. While our numerical results point toward a phase-matching synchronization mechanism, a definitive analytical proof of the SIFF effect is still required to formally evaluate the existence range of these solutions. Furthermore, the sensitivity of the effect to different backreflection strengths and its impact on the overall power conversion efficiency, whether SIFF improves or degrades the forward comb flux compared to regular PhCRs, merit dedicated investigation.

In addition, our SIFF method elegantly resolves the fundamental incompatibility between SIL and normal-dispersion PhCRs: by enforcing forward propagation, it enables self-injection-locked platicon microcombs that combine the stability advantages of SIL with the unique spectral properties of normal-dispersion combs.
By providing a scalable path to forward comb operation in normal dispersion PhCR, the SIFF effect represents an important building block for the next generation of robust, fully integrated normal-dispersion microcomb sources.

\section*{Acknowledgments}
E.L. acknowledges support from the Swiss National Science Foundation (SNSF) under contract no. 191705. 
This project has received funding from the European Union's Horizon Europe research and innovation programme under grant agreement No 101137000.

We acknowledge Travis Briles and David Carlson for assistance fabricating and characterizing the devices, as well as fruitful discussion with Tobias Herr.

\section*{Competing Interests}
The authors declare no competing interests.
This work was supported by NIST, and this paper is not subject to copyright in the US. Trade names are provided for information only and do not represent an endorsement by NIST.

\printbibliography
\end{refsection}

\clearpage

\appendix
\section*{Supplementary Material}
\setcounter{subsection}{0}
\setcounter{figure}{0}
\renewcommand\thefigure{S\arabic{figure}}
\makeatletter
\renewcommand{\fnum@figure}{\textbf{Supplementary Figure~\thefigure}}
\makeatother
\begin{refsection}
\subsection{Impact of parasitic back-reflections on SIFF dynamics}

\begin{figure*}
	\centering%
	\includegraphics[width=\textwidth, height=0.8\textheight, keepaspectratio]{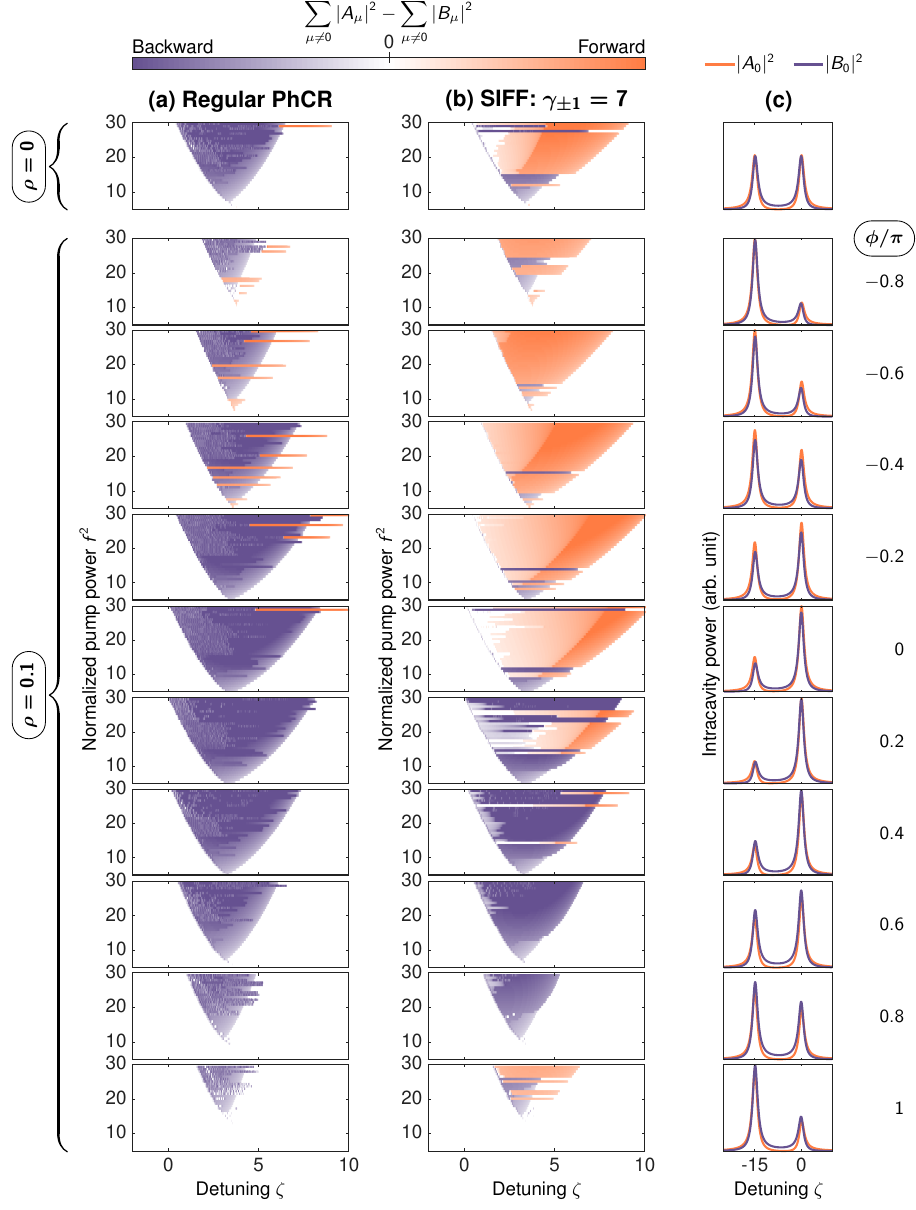}%
	\addsubcap{fig:SIReflct:Standard}{Simulated comb `phase diagrams' showing the comb main direction in a regular PhCR ($\gamma_0 = 15$, $\gamma_{\pm 1} = 0.7$). The case compare the absence of reflection ($\rho=0$) and \qty{10}{\percent} reflection with varying reflection phase.}%
	\addsubcap{fig:SIReflct:Sim:SIFF}{Same analysis for a SIFF case, where $\gamma_0 = 15$ and $\gamma_{\pm 1} = 7$}%
	\addsubcap{fig:SIReflct:Sim:split}{Representation of the splitting of mode $\mu=0$ at low power, accounting for the backreflection and phase.}%
	\caption{%
		\textbf{Backreflection effect}
		\processCaptions
	}
	\label{fig:SIReflct}
\end{figure*}

While the main text assumes an idealized case ($\rho = 0$), we performed additional simulations to evaluate the robustness of the SIFF effect under realistic experimental conditions, incorporating a back-reflection of $\rho = \qty{10}{\percent}$. We compared a regular PhCR ($\gamma_0 = 15, \gamma_{\pm 1} = 0.7$) with a SIFF-enabled device ($\gamma_{\pm 1} = 7$) across various back-reflection phases $\phi \in [-\pi, \pi]$. The resulting phase diagrams, comprising simulated detuning scans ($\zeta$) over varying pump powers ($f^2$), are summarized in \cref{fig:SIReflct}.

Our analysis shows that the SIFF effect is globally preserved even when accounting for back-reflections; the SIFF-engineered device consistently displays a broader and more reliable forward-propagating operation range compared to the regular PhCR. However, the dynamics are sensitive to the reflection phase $\phi$, which dictates the power distribution between the red- and blue-shifted supermodes. As evidenced in \cref{fig:SIReflct:Sim:split}, certain phases lead to a significantly higher comb generation threshold when the power coupled into the red-shifted mode (at $\zeta \approx 0$) is minimized.

Moreover, the phase $\phi$ directly modifies the forward-backward power balance at the pump resonance. For instance, at $\phi = -0.4\pi$, the forward power dominates at $\zeta=0$, facilitating an efficient SIFF transition. Conversely, at $\phi = 0.6\pi$, the backward power becomes dominant in the pumped mode, effectively canceling the SIFF effect and restoring backward comb operation.
In such cases, the side-mode coupling $\gamma_{\pm 1}$ may become suboptimal to override the enhanced backward bias induced by the reflection.

While a comprehensive optimization for every reflection phase falls beyond the scope of this study, our analysis underscores the critical role of back-reflections in defining microcomb thresholds and dynamics, echoing recent literature~\cite{Zang2025LaserPower}, even at the relatively low levels (10\%) observed here.
Crucially, such residual reflections are not fundamental limits but engineering artifacts that can be mitigated, or even strategically utilized through proper device design, such as optimized angled or coated facet, and integrated phase shifters.
This highlights a promising path to further enhance the SIFF effect by co-engineering the PhC parameters and the global cavity feedback to achieve a synergistic enhancement in integrated directional control.

\subsection{Robustness of SIFF with dispersion}
\begin{figure*}
	\centering%
	\includegraphics[width=\textwidth, height=0.8\textheight, keepaspectratio]{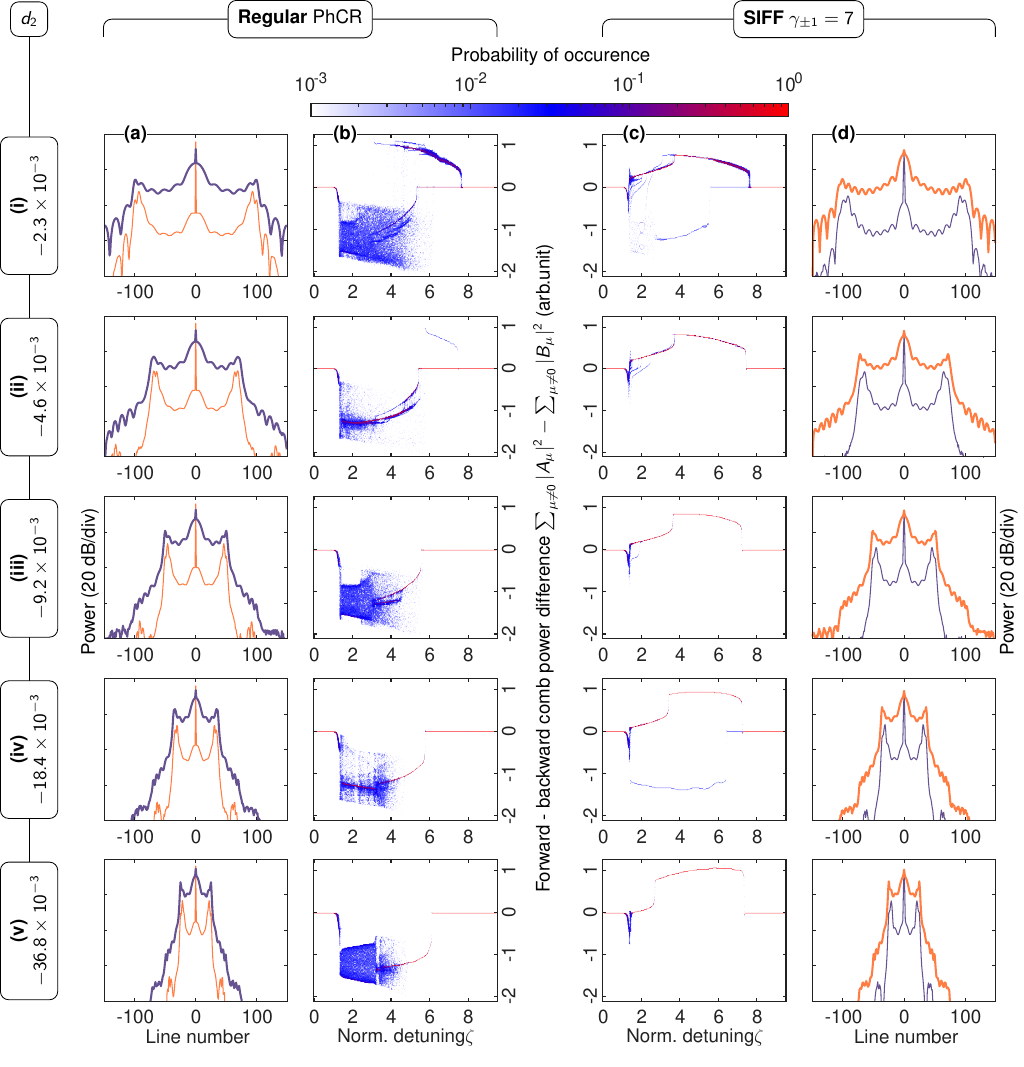}%
	\caption{%
		\textbf{Robustness of SIFF across dispersion regimes.} \textbf{(a,b)} regular PhCR and \textbf{(c,d)} SIFF device for dispersion parameters $d_2$: \textbf{(i)} -0.0023, \textbf{(ii)} -0.0046, \textbf{(iii)} -0.0092, \textbf{(iv)} -0.0184, (v) -0.0368. \textbf{(a,d)} Most probable spectra at detuning $\zeta=5$. \textbf{(b,c)} Persistence maps computed over 50 detuning scans. SIFF effect is retained across all dispersion cases.
	}
	\label{fig:SIDisp}
\end{figure*}

To investigate the dependence of SIFF on cavity dispersion, we performed detuning scans for different dispersion parameters $d_2$ at a fixed pump power $f^2=20$. \Cref{fig:SIDisp} shows the results for $d_2$ values spanning more than one decade: \numlist[exponent-mode=engineering, list-exponents=combine-bracket]{-0.0023; -0.0046; -0.0092; -0.0184; -0.0368} (rows i-v). The first two columns (a,b) correspond to a regular PhCR ($\gamma_0 = 16$, $\gamma_{\pm 1} = \bar{\gamma} = 0.88$), while the last two columns (c,d) correspond to the SIFF case ($\gamma_0 = 16$, $\gamma_{\pm 1} = 7$).
The middle columns (b,c) show ``persistence maps'',i.e. probability distribution maps of the forward-backward power difference during detuning scans, computed over 50 scans, with color scale indicating probability density. These demonstrate that the SIFF effect is retained throughout all dispersion cases: the SIFF device produces near-systematic forward combs across the entire range.
In contrast, for the regular PhCR, lower normal dispersion induces more regular switching and more instabilities than stronger dispersion, where the combs are systematically backward.

The outermost columns (a,d) show the most probable spectra at detuning $\zeta=5$. As dispersion strength increases, the spectral extension shrinks in both cases, with dispersive features moving closer to the pump. In the standard device (a), combs are backward and modulated, indicating dual-pulse formation. In the SIFF device (d), combs show forward-dominant power and smoother spectral envelopes associated with single-pulse formation.

These results confirm that the optimal SIFF setting is retained over more than one decade of dispersion change, demonstrating robustness against cavity dispersion variations.

\printbibliography

\end{refsection}

\vfill\null
\clearpage

\end{document}

%% file: headerCommands.tex
\usepackage[hmargin=1.9cm,vmargin=2.5cm]{geometry}
 
\usepackage[utf8]{inputenc}
\usepackage[T1]{fontenc}
\usepackage[english]{babel}
\usepackage{hyperref}
\usepackage{url}

\usepackage{changes}
\usepackage{lipsum}
\setcommentmarkup{\footnotesize\itshape\color{red}-- #1 --}

\usepackage{amsmath}
\usepackage{amsfonts}
\usepackage{amssymb}
\usepackage{mathtools}
\usepackage{lmodern}
\usepackage{upgreek}
\usepackage{ragged2e}
\usepackage{cuted}
\usepackage[version=3]{mhchem}
\usepackage{siunitx}
\usepackage{circledsteps}

\usepackage{titling}
\usepackage{titlesec}
\usepackage[affil-it]{authblk}

\pretitle{\begin{center}\large\bfseries}
	\posttitle{\par\end{center}}
\preauthor{\begin{center}
		\lineskip 0.25em%
		\begin{tabular}[t]{c}}
		\postauthor{\end{tabular}\par\end{center}}
\predate{\begin{center}\small}
	\postdate{\par\end{center}\vskip-1cm}
\date{}

\titleformat{\section}{\normalfont\large\bfseries}{}{0pt}{}
\titlespacing{\section}{0pt}{5pt plus 5pt minus 3pt}{2pt}

\titleformat{\subsection}[runin]{\normalfont\bfseries}{}{0pt}{}[\mbox{ -- }]
\titlespacing{\subsection}{0pt}{4pt}{4pt plus 2pt minus 1pt}

\titleformat{\subsubsection}[runin]{\normalfont\itshape}{}{0pt}{}[: ]
\titlespacing{\subsubsection}{0pt}{3pt minus 3pt}{4pt plus 2pt minus 1pt}

\makeatletter
\renewcommand{\fnum@figure}{\textbf{Figure~\thefigure}}
\makeatother

\usepackage[%
backend=biber,
style=nature,
alldates=year,
doi=false,
isbn=false,
url=false,
eprint=true
]{biblatex}

\DeclareSourcemap{
	\maps{
		\map{
			\step[fieldset=note, null]
			\step[fieldset=address, null]
			\step[fieldset=location, null]
			\step[fieldset=language, null]
		}
	}
}

\newbibmacro{string+doiurl}[1]{%
	\iffieldundef{doi}
	{\iffieldundef{url}
		{#1}
		{\href{\thefield{url}}{#1}}}
	{\href{https://doi.org/\thefield{doi}}{#1}}}

\DeclareFieldFormat{title}{\usebibmacro{string+doiurl}{\mkbibemph{#1}}}

\usepackage{enumitem}
\usepackage{placeins}

\usepackage{cleveref}

\usepackage{subcaption}

\DeclareCaptionLabelFormat{paren}{(#2)}
\DeclareCaptionJustification{justified}{\justifying}
\newcommand{\figsublabel}[1]{\protect\phantomsubcaption\label{#1}\protect\bfseries\textsf{\subref*{#1}}}%

\usepackage{xparse} %
\usepackage{xspace}

\ExplSyntaxOn
\prop_new:N \l_pollard_a_prop
\cs_new:Npn \pollard_wrappera:nn #1#2
{%
	\textbf{\subref*{#1}}\nobreakspace\ignorespaces#2\unskip\ignorespacesafterend\hspace{4pt plus 2pt minus 2pt}%
}

\NewDocumentCommand { \addsubcapLbl } {sO{black}m m m m }
{%
	\IfBooleanTF#1
	{\def \fillingArg{}}
	{\def \fillingArg{fill=white}}
	\protect\node[rectangle, text=#2, anchor=north~west,style/.expanded=\fillingArg] at (#4,#5) {\figsublabel{#3}};
	\prop_gput:Nnn \l_pollard_a_prop {#3} {#6}
}
\NewDocumentCommand { \addsubcap } { m m }
{%
	{\phantomsubcaption\label{#1}}%
	\prop_put:Nnn \l_pollard_a_prop { #1 } { #2 }%
}

\NewDocumentCommand { \processCaptions } { }
{
	\prop_map_function:NN \l_pollard_a_prop \pollard_wrappera:nn
	\prop_gclear_new:N  \l_pollard_a_prop
}
\ExplSyntaxOff

\usepackage{epstopdf}
\usepackage{graphicx}
\graphicspath{{./Figures/}}

\usepackage{varwidth}
\usepackage{xcolor}
\definecolor{mossgreen}{rgb}{0.68, 0.87, 0.68}
\definecolor{mustard}{rgb}{1.0, 0.86, 0.35}
\definecolor{frenchblue}{rgb}{0.0, 0.45, 0.73}
\definecolor{lightskyblue}{rgb}{0.53, 0.81, 0.98}
\definecolor{purpleheart}{rgb}{0.41, 0.21, 0.61}
\definecolor{jasper}{rgb}{0.84, 0.23, 0.24}

\usepackage{tikz}
\usetikzlibrary{calc,fit,shapes,arrows,positioning,shadows,shapes.symbols}

\tikzset{block/.style = {
		rectangle, draw, fill=frenchblue!25, rounded corners, minimum height=3em,inner sep=5pt,%
		drop shadow,
		execute at begin node={\begin{varwidth}{#1}\centering},%
			execute at end node={\end{varwidth}}
	},
	block/.default={9em}
}
\tikzstyle{decision} = [diamond, aspect=1.5, draw, fill=jasper!30, 
drop shadow,
text width=4em, text badly centered, inner sep=0pt,outer sep = -.5pt]
\tikzstyle{cloud} = [draw, ellipse,fill=mustard,
drop shadow,
minimum height=2em]
\tikzstyle{line} = [draw, -stealth, line width=1pt]

%% file: authors.tex
\author[1]{Erwan Lucas\thanks{erwan.lucas@polytechnique.org}}

\author[2,3,4]{Su-Peng Yu}

\author[2,3]{Jizhao Zang}

\author[2,3]{Scott~B.~Papp}

\affil[1]{Laboratoire ICB, UMR 6303 CNRS Université de Bourgogne, 21078 Dijon, France}
\affil[2]{Time and Frequency Division, National Institute of Standards and Technology, Boulder, CO USA}
\affil[3]{Department of Physics, University of Colorado, Boulder, CO 80309, USA}

\affil[4]{Current affiliation: Quantum Valley Ideas Laboratories, Ontario N2L 0A7, Canada}